\documentclass[12pt]{article}

\usepackage{booktabs}
\usepackage{longtable}
\usepackage{siunitx}
\usepackage{amsmath}
\usepackage[utf8]{inputenc}
\usepackage{amssymb}
\usepackage{authblk}%
\usepackage[numbers]{natbib} 
\setcitestyle{square}
\usepackage{graphicx} 
\usepackage{amsmath}
\usepackage{hyperref}
\usepackage{longtable}
\usepackage{float}
\usepackage{subcaption}
\usepackage{booktabs} 
\usepackage{newunicodechar}
\newunicodechar{₁}{$_1$}
\newunicodechar{₀}{$_0$}
\usepackage{textcomp}
\usepackage{newunicodechar}
\newunicodechar{≈}{\approx}
\newunicodechar{−}{\ensuremath{-}}
\usepackage{textcomp} 
\usepackage{verbatim,comment} 
\usepackage{caption}
\usepackage{textcomp}
\usepackage{booktabs}
\usepackage{setspace}
\usepackage{graphicx}
\usepackage{longtable}
\usepackage[export]{adjustbox}

\usepackage[
    left=0.75in,
    right=0.75in,
    top=0.8in,
    bottom=0.8in
]{geometry}

\usepackage[
justification=raggedright,
singlelinecheck=false,
font=small
]{caption}

\title{\vspace{-1cm}Patterns in Individual Blood Count Trajectories in the UK Biobank Characterise Disease-Specific Signatures and May Anticipate Pan-Cancer Risk}
\author[1,2]{Riya Nagar}
\author[2,3]{Abicumaran Uthamacumaran}
\author[1]{Adelaide de Vecchi}
\author[1,2,4,5]{Hector Zenil\thanks{Corresponding author: hector.zenil@algocyte.ai, hector.zenil@kcl.ac.uk}}
\affil[1]{\normalsize\text{ }Research Departments of Biomedical Computing and Digital Twins, School of Biomedical Engineering and Imaging Sciences, King's College London, U.K.}
\affil[2]{\normalsize\text{ }Oxford Immune Algorithmics, Oxford University Innovation and London Institute of Healthcare Engineering, London, U.K.}
\affil[3]{\normalsize\text{ }McGill University, Department of Surgical and Interventional Sciences and Neurosurgical Simulation and AI Learning Centre, Montreal, Canada. \small}

\affil[4]{\normalsize\text{ }British Society for Research on Ageing, U.K.}
\affil[5]{\normalsize\text{ }Algorithmic Dynamics Lab, King's Institute for Artificial Intelligence, King's College London, U.K.}

\date{ }

\begin{document}
\doublespacing
\maketitle

\begin{abstract}
Routine blood tests are typically interpreted as individual measurements against population-derived reference ranges, with limited use of information contained in changes across repeated tests. Here, we investigate whether trends in routinely measured haematological markers can reveal patterns associated with ageing and disease progression across cancers, cardiovascular diseases and infections. Using retrospective Complete Blood Count (CBC) data from the UK Biobank, we examine changes in individual biomarkers and their relationships across successive measurements, while assessing potential confounding factors. We combine statistical analysis, machine learning and network-based approaches to identify patterns associated with different disease states and stages. We find that clinically relevant information is contained not only in absolute blood-cell counts, but also in the direction, magnitude and coordinated patterns of change between measurements. These patterns provide candidate disease signatures that may complement conventional interpretation based on individual values and reference intervals. Because the approach relies on inexpensive tests already collected routinely in clinical practice, it offers a practical route for extracting additional clinical information from existing data, with potential applications in earlier disease detection, monitoring and precision healthcare.

\end{abstract}

\section{Introduction}

Traditionally, a blood test such as the Complete Blood Count (CBC) is a first-line of defense and investigation to a wide range of possible clinical presentations. Together with medical imaging, blood testing is a cornerstone of modern diagnostic medicine \cite{seo2022}. Because blood perfuses every organ in the body \cite{matienzo2020anatomy}, is the main transportation highway of nutrients, and is the communication channel of the immune system, changes in its constituents can signal early deviations from normality across a wide range of conditions and pathologies \cite{hernandez2026integrative,virdee2024}. Yet the diagnostic potential of routine haematological markers remains largerly underused specially to build individual's baselines and understand temporal trends~\cite{foy2025haematological}: they are typically interpreted through binary normal v abnormal decisions based on population reference intervals applied at a single timepoint, in response to specific symptoms \cite{timbrell2024, heydari2024}, rather than as part of an individual’s temporal trajectory in their transition between health and disease. This distinction matters most for two groups of patients: those in whom a trajectory precedes diagnosis (incident cases), where it offers a window onto early shifts in blood markers, and those in whom it reflects already established disease (prevalent cases), where it can instead support ongoing disease monitoring and  potentially alongside a hypothesised, treatment-altered haematological profile \cite{ferle2024, foy2025haematological}. Beyond this, discovering the hidden information contained in routine longitudinal blood tests can unlock a radical change in our understanding of health as a static state that disease can supersede, providing a nuanced assessment of healthy individuals and homeostatic balance \cite{hernandez2026integrative, foy2025haematological,herold2022} and offers a potential poweeful tool for precision healthcare and predictive medicine. Despite this potential, no validated or scalable framework yet translates routine longitudinal blood testing into a dynamic, disease-agnostic measure for risk assesment deployable at population scale across multiple disease types \cite{cohen2021personalized}.

A key advantage of this approach is that it leverages inexpensive, globally standardised tests already conducted at mass scale, unlike design-for-purpose tests that are difficult to scale for population-wide use. The vast archive of longitudinal blood tests held in hospital databases is therefore a unique opportunity to study intra- and inter-individual variation in blood biomarkers across health, disease, and ageing \cite{pyrkov2021}; a necessary step towards dynamic, population-scale risk prediction and a shift from reactive to proactive healthcare. Dynamical systems methods in particular remain underused in this space, likely due to a scarcity of longitudinal data \cite{uthamacumaran2022review}. Here, we exploit this under-used longitudinal resource to advance precision healthcare and predictive medicine across a range of common diseases.

As reference normal-abnormal intervals can vary widely between individuals, populations and legacy systems adopted across countries\cite{refvalues,wu2025similarity}, they lack the granularity necessary to capture potential health risks in a timely manner. Multiple studies have shown that intra-subject variations in CBC biomarkers over short periods of time are generally narrower than population-based reference ranges\cite{foy2025haematological}, even in the presence of acute environmental changes such as space flight~\cite{aziz2019biological}. Recent work showed that the homeostatic values of most CBC markers are remarkably stable for each individual over periods of up to 20 years, while their inter-individual variation within the reference intervals can be large~\cite{foy2025haematological}. This evidence points to the existence of healthy mean values of blood biomarkers that are specific to the individual. Models of dynamic risk prediction and immunological state should therefore be based on the deviation from these more granular reference values, rather than a single population-wide baseline, over a period of time.
 
Resources such as the UK Biobank\cite{biobank1,biobank2, sudlow2015}now make it possible to study haematological trajectories at population scale. Yet despite growing evidence for stable, individual-specific haematological baselines \cite{foy2025haematological}, no study to date has combined longitudinal CBC trajectories with a single, explainable, disease-agnostic risk index validated prospectively across major disease catagories including cancer, cardiovascular disease and chronic infection \cite{tanriver2023additive}. Additionally, in a previous retrospective study, we introduced a Numerical Immune Score (NIS) that aggregates routine blood markers into a single explainable metric of health status related to biological age \cite{hernandez2026integrative}, but this relied on single-timepoint measurements across the full CBC feature space. Here we extend that work along three key methodological axes: (A) moving from single-timepoint to longitudinal tracking of the index; (B) testing whether restricting the index to a smaller, disease-specific subset of analytes sharpens early detection ahead of diagnosis, rather than relying on the full CBC panel; and (C) moving from a single disease-agnostic score to a granular, disease-specific approach, descending from broad disease categories (cancer, cardiovascular disease, chronic infection) down to individual cancer subtypes.

Here, we address this gap by extending our indexes to longitudinal UK Biobank data, assessing the early detection and predictive value of within-individual trends across solid and blood cancers, cardiovascular disease and chronic infection. We further examine whether restricting the index to a smaller, disease-specific subset of significant analytes, rather than the full CBC panel, can sharpen discrimination between disease and control trajectories. Our results suggest that longitudinal analysis of inexpensive, widely available blood tests may improve prediction and classification of a range of complex diseases relative to single-timepoint measures, contributing to closing the current gap in validated, scalable tools for translating routine blood data into dynamic, prospective risk assessment. This may have implications beyond detection: real-time disease-signature discovery could support prognostic evaluation and preventive care, and point towards therapeutic targets for precision medicine, without requiring additional disease-specific tests. 

\section{Results}

\subsection{Pan-disease haematological patterns}

Diseases were classified as incident, diagnosed by ICD-10 code after the first blood test (time point 0), or prevalent, diagnosed prior to study admission \autoref{fig:UKBBcohort}. Between-group pairwise comparisons of complete blood count (CBC) markers, following significant Kruskal–Wallis and Bonferroni-corrected Mann–Whitney U tests (p$<$0.05), revealed widespread differences across disease groups, in both incident and prevalent categories (\autoref{tab:S4_CBC_only_summary}).

At initial assessment, 15-16 analytes differed significantly per pairwise comparison, with the largest contrasts for CVD-vs-control and infection-vs-control (\autoref{tab:S2_top_effects}, \autoref{tab:S4_CBC_only_summary}); by first follow-up, significant differences remained but were fewer, with the strongest separation between cancer and Rest (15 analytes) and CVD and Rest (14 analytes). By contrast, the magnitude of longitudinal change (delta) differed significantly between groups for only 4-5 analytes.

Within-group longitudinal analysis (Friedman test, p$<$0.05) identified widespread analyte–timepoint changes: 19 analytes in cancer, 17 in CVD, and 11 in chronic infection (\autoref{tab:S5_within_group_summary}), concentrated in MCH, MCHC, MCV, platelet count, basophil count, and neutrophil percentage (\autoref{tab:S6_top_longitudinal_effects}, \autoref{tab:S7_targeted_longitudinal}); red cell indices and platelet-related measures are of particular interest, having previously been linked to survival and disease burden in cancer and CVD, as expanded on in the discussion. This makes their longitudinal divergence a plausible early signal rather than incidental variation. These reflect the trend direction criterion defined in Methods: in CVD, MCH, MCHC, monocyte percentage, and platelet count increased over time while RBC, neutrophil count, and haematocrit decreased; cancer showed comparable increases in MCH and MCHC alongside decreases in neutrophil percentage and monocyte count, consistent with the myeloid-skewed profile discussed in the Discussion. Chronic infection remained comparatively stable.

As illustrated in \autoref{fig:cvd,cancer,infection}, we identified distinct trajectories for specific analytes, including MCH, lymphocyte count, and monocyte count (\autoref{tab:S3_targeted_analytes}), across disease groups using normalised longitudinal plots, many without overlapping IQR bands. LME models across the three major disease groups also identified several significant analyte changes over time \autoref{fig:cvd,cancer,infection}. In cancer, red blood cell indices — including MCH concentration, MCV, RBC counts, and haemoglobin — showed interaction significance alongside some immune cells. For CVD, interactions were weaker but still significant for some analytes, including red blood cell indices such as MCH and MCV and white blood cell subtypes such as lymphocyte and neutrophil percentage. LME modelling identified only two analytes significantly associated with chronic infection: monocyte count and monocyte percentage.
\begin{figure}[!ht]
\centering
\includegraphics[width=1\textwidth, page=3]{Biobank_global.pdf}
\caption{(a) Min–max normalised trajectories of CBC analytes across three UK Biobank blood collection waves (2006--2010, 2011--2012, and 2014+). Panels show MCH, lymphocyte count, and monocyte count. Lines indicate median trajectories for Rest (green), cancer (red), CVD (blue), and infection (orange), with shaded bands representing the interquartile range (IQR) of the Rest group. Grey circles mark the mean diagnosis year for each disease group. The separation and divergence in trend before the grey dots are the opening of windows of opportunity for earlier detection. (b) A Heatmap showing the statistical significance of time $\times$ disease group interactions from linear mixed-effects models comparing cancer, CVD, and chronic infection with the Rest reference group. Rows represent CBC analytes and columns represent disease groups. Colours show $-\log_{10}$ of FDR-adjusted p-values (Benjamini-Hochberg), with black dots indicating analytes significant after FDR correction (FDR $\leq 0.05$).}
\label{fig:cvd,cancer,infection}
\end{figure}

\subsection{Cancer-specific haematological signatures}
Within incident cancer cases, our results may reflect the transition from healthy to malignancy, while in prevalent cases, they may indicate persistently altered haematological profiles, as suggested by the potential-max normalised plots in \autoref{fig:cancerIP.png}. 

\subsubsection{Global cancer signals}

Across CBC markers, significant differences were observed between Rest, prevalent cancer, and incident cancer groups, both cross-sectionally and longitudinally. Between-group analysis identified 13 significant markers, with the strongest effects in RBC, haemoglobin, lymphocyte measures, and haematocrit, primarily distinguishing Rest from both cancer groups and, to a lesser extent, prevalent from incident cancer. Longitudinally, widespread temporal changes were observed, with 18 markers in incident cancer and 15 in prevalent cancer showing at least one significant shift. The strongest signals across all groups were in red cell indices (MCH, MCHC) and platelet markers (platelet count, mean platelet volume), alongside monocyte and neutrophil proportions.

\begin{figure}[ht]
\centering
\includegraphics[width=\textwidth]{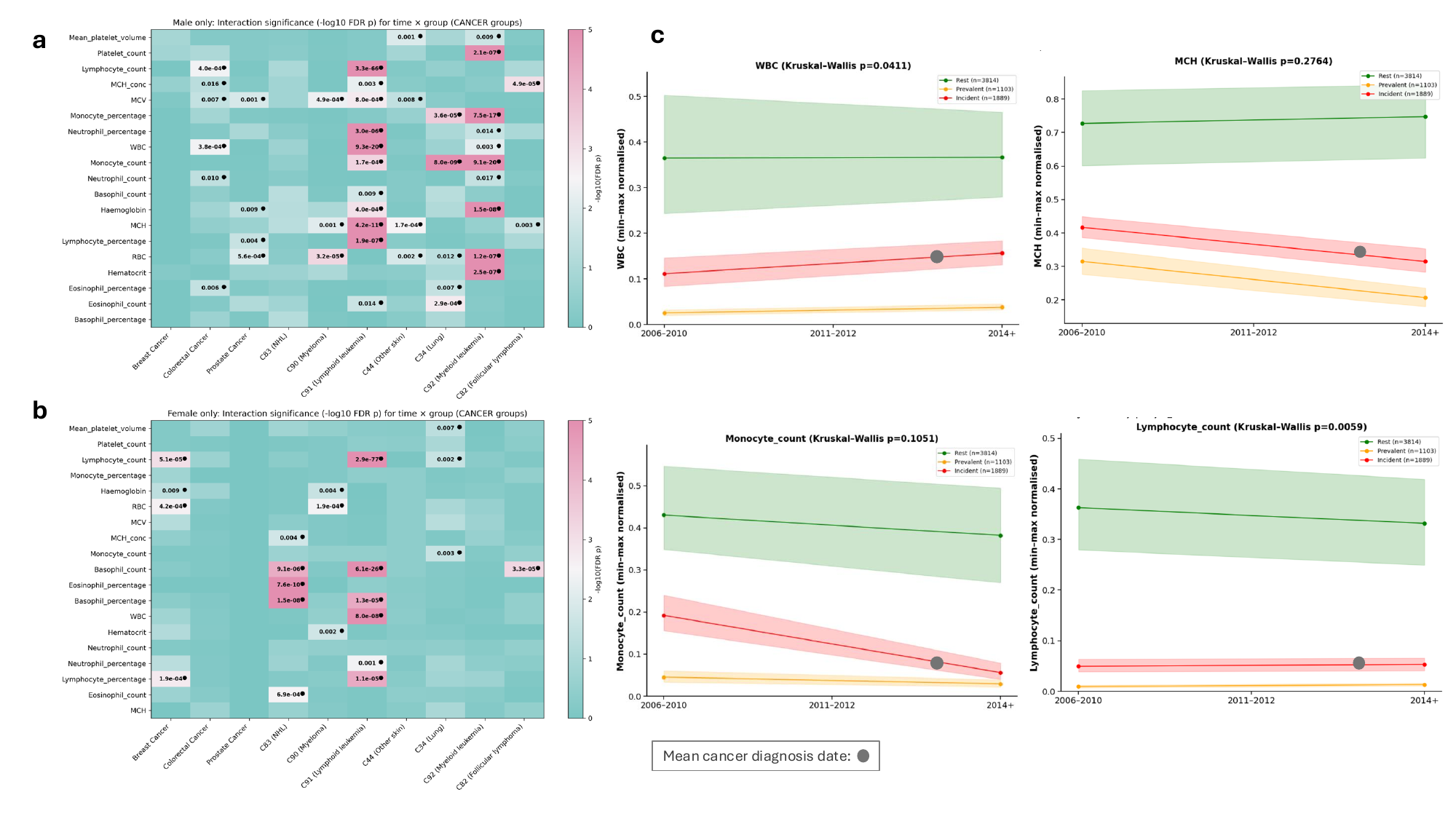}
\caption{(a–b) Heatmaps showing interaction significance (−log10 FDR-adjusted p-values) from linear mixed-effects (LME) models for the effect of time × cancer group across CBC analytes, stratified by sex: (a) males and (b) females. ignificant analytes are marked with a black circle and labelled with their p-value. (c) Normalised longitudinal trajectories of selected CBC markers (WBC, MCH and monocyte count) across Rest, incident cancer, and prevalent cancer groups. Shaded areas represent interquartile range (IQR) and the grey marker denotes the mean cancer diagnosis time point for incident cases.}
\label{fig:cancerIP.png}
\end{figure}

White blood cell subtypes showed distinct trajectories across groups, reflecting the transition from health to malignancy in incident cancer versus established disease and recovery in prevalent cancer. Monocyte counts were consistently higher in both cancer groups than Rest, while lymphocyte levels declined progressively from Rest to incident cancer, indicating a shift toward a more myeloid-dominant profile. Within-group trends further showed monocytes decreasing longitudinally in both prevalent and incident cancer, and MCH declining over time with moderate effect sizes. These patterns were supported by non-overlapping IQR bands in normalised trajectories, as seen in \autoref{fig:cancerIP.png}, suggesting distinct temporal profiles. Overall, incident cancer showed the strongest, most consistent temporal shifts, particularly in platelet and neutrophil markers, while prevalent cancer showed intermediate patterns. As our design provides only two informative timepoints per individual, we cannot formally disentangle disease progression from treatment effects or recovery using the UK Biobank data; these intermediate patterns are consistent with an altered haematological profile broadly reflecting established disease, rather than distinguishing between these processes.

\subsubsection{Variations between blood and solid cancers}

To provide subgroup-level insight, we stratified cancer cases into blood cancer and solid tumours. Between-group analysis (Kruskal–Wallis with Bonferroni-corrected Mann–Whitney U tests) identified 8 significant CBC markers, with 13 significant pairwise contrasts overall. The strongest effects were in erythroid markers, particularly haemoglobin, haematocrit, and RBC, which were significantly lower in blood cancer than both Rest and solid tumours, with moderate effect sizes. These markers also showed smaller but significant reductions in solid tumours relative to Rest, indicating a gradient of erythroid suppression. Differences between Rest and solid tumours were more modest and involved additional markers, including lymphocyte count and percentage, MCV, MCHC, and basophil count. Longitudinally, both Rest and solid tumour groups showed widespread temporal changes, strongest in red cell indices (MCH, MCHC) and platelet markers, alongside neutrophil and monocyte proportions. Solid tumours showed more pronounced immune and platelet-related dynamics over time. These findings point to distinct profiles; blood cancer marked by erythroid suppression, solid tumours by moderate, multi-lineage and temporal immune shifts. 

\subsubsection{Distinct disease signatures for different cancer types}

To further granularity, incident cancer cases were further stratified into the ten most common cancer types within the UK Biobank as seen in \autoref{fig:UKBBcohort}. Linear mixed-effects (LME) models identified sex and disease specific analyte interactions over time, as illustrated in the heatmaps in \autoref{fig:cancerIP.png}.

In solid tumour cancers, significant interactions were driven primarily by red blood cell indices (MCV, RBC, haemoglobin) and immune cell proportions (lymphocyte and monocyte percentages, eosinophil count), with RBC and haemoglobin showing the strongest effects. Blood cancers were characterised by lymphocyte-related changes, with lymphocyte count showing the highest interaction significance, followed by erythroid markers (haemoglobin, haematocrit, RBC) and platelet measures (platelet count, mean platelet volume). These patterns indicate distinct signatures across cancer types, such as  combined erythroid and immune shifts in solid tumours versus lymphocyte driven dynamics in blood cancers, with specific analyte combinations varying by malignancy and sex.

\subsubsection{Cancer-related immune score}

\begin{figure}[H]
\centering
\includegraphics[width=\textwidth]{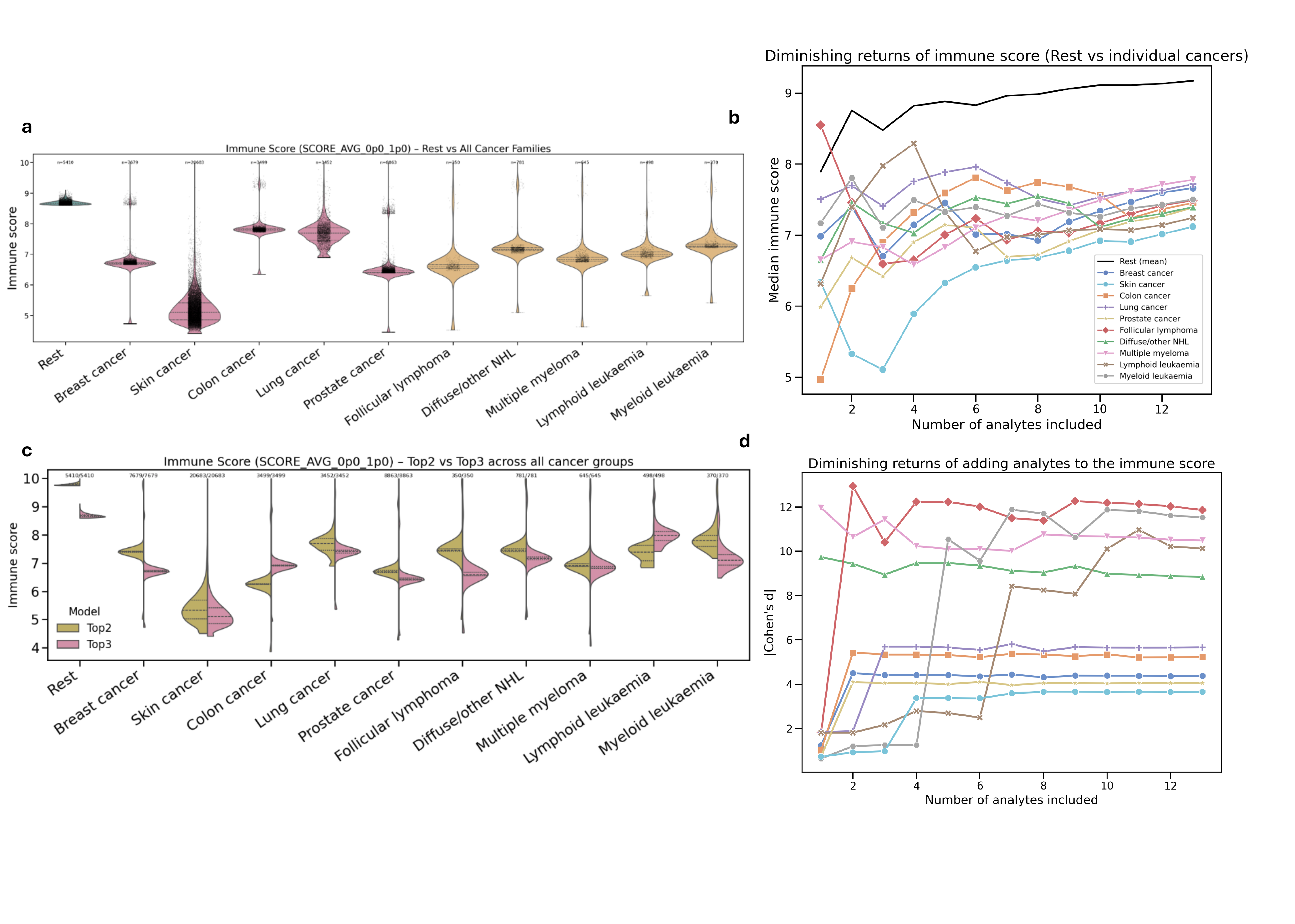}
\caption{From feature selection to early detection and prediction in incident cancer patients. Blood analytes were ranked using a linear mixed-effects (LME) model applied to longitudinal blood measurements and sequentially added to an immune score. (a) Distribution of immune scores across cancer groups compared with the Rest group when using all significant analytes highlighted in the LME models.(b) Effect size (Cohen's d) between each cancer group and the Rest group as increasing numbers of analytes are included, showing diminishing returns after the first few markers. (c) Immune score distributions across cancer groups comparing scores constructed using the top two versus the top three ranked analytes. (d) Mean immune score for the Rest group and individual cancer types as a function of the number of analytes included.   }
\label{fig:diminishingreturns}
\end{figure}

We evaluated how significant LME-identified markers contribute to immune score construction and group discrimination, adding analytes sequentially by effect size. As \autoref{fig:diminishingreturns} (d) shows, the largest separation between cancer groups and Rest (Cohen's d) is achieved with the first few highest-ranked analytes for most malignancies, with subsequent markers adding minimal value; in several cases performance plateaus or slightly decreases, indicating the discriminative signal is concentrated in a small marker subset. Similarly, \autoref{fig:diminishingreturns} (b) shows the median immune score reaching peak divergence between Rest and cancer groups with few analytes, with additional markers sometimes reducing separation clarity, likely reflecting analyte redundancy. Together, these results indicate a compact, high-ranking marker subset captures most of the predictive signal, with additional markers offering limited or confounding contribution.

We further compared immune scores computed from all significant analytes against this reduced-marker approach. Using all significant analytes per malignancy (\autoref{fig:cancerIP.png}), immune score distributions showed clear separation between Rest and cancer groups, and between cancer groups, despite variation in the analyte combinations driving each disease. Notably, comparing immune scores from the top two versus top three ranked analytes (Figure 4d), the top two already achieved strong separation, with only marginal improvement from a third marker, and sometimes reduced performance. These findings build on our previous work \cite{hernandez2026integrative}, which used the full CBC panel, by showing that a reduced, disease-specific analyte subset retains substantial predictive power,  suggesting a small number of key haematological markers suffice to capture meaningful biological differences between healthy individuals and specific cancer types, enabling more targeted, interpretable risk stratification.

\subsection{Haematological patterns in non-cancer conditions}

\subsubsection{Cardiovascular disease (CVD)}

CVD was defined broadly to encompass chronic vascular diseases, including high blood pressure (HBP), angina, heart attack, and stroke, with individuals classified as prevalent (diagnosed before recruitment) or incident (diagnosed after). Further stratification was limited by co-morbidities within the CVD classification. Across CBC markers, significant differences were observed between Rest, incident CVD, and prevalent CVD, both cross-sectionally and longitudinally. Between-group differences were driven primarily by erythroid markers (RBC, haemoglobin, haematocrit), platelet-related measures, and immune cell proportions (neutrophil, monocyte percentages), with the strongest effects between Rest and prevalent CVD and more modest shifts in incident CVD. Longitudinally, the most consistent signals appeared in red cell indices (MCH, MCHC), platelet-related markers, and neutrophil and monocyte proportions, with prevalent CVD showing the strongest and most consistent temporal shifts. These patterns are reflected in the normalised trajectories \autoref{fig:CVD}: MCH declines over time in prevalent CVD versus relatively stable trends in Rest, mean platelet volume rises progressively across CVD groups, and RBC remains comparatively stable with only subtle group separation. Incident CVD consistently falls between Rest and prevalent CVD. Together, these findings highlight progressive haematological alterations from health to disease, greatest in established CVD. When stratified by sex in LME modelling, males showed a broader set of significant analytes than females, with erythroid markers (e.g. MCV, MCH, haemoglobin) predominantly significant in males, fewer significant analytes in females, and limited overlap between sexes \autoref{fig:CVD}. 

\begin{figure}[H]
\centering
\includegraphics[width=\textwidth, page=7]{cvd.pdf}
\caption{(a) Median min–max normalised trajectories of MCH, WBC, and Mean Platelet Volume across three timepoints by CVD status (Rest, Prevalent, Incident), with IQR shading; grey dots indicate average diagnosis timing in the incident cases. (b) Heatmap of sex-stratified time $\times$ CVD interaction effects from linear mixed-effects (LME) models, shown as $-\log_{10}$ FDR-adjusted p-values. Rows represent CBC analytes and columns represent CVD\_Male and CVD\_Female groups relative to the Rest reference. Black dots indicate analytes that remain significant after FDR correction (q $<$ 0.05).}
\label{fig:CVD}
\end{figure}

\subsubsection{Chronic infection}

Chronic infection, defined as persisting for longer than 6 months, corresponded to ICD10 chapter 1 (certain infectious and parasitic diseases). Conditions included prevalent tuberculosis (TB), HIV, viral hepatitis, herpes virus, and sequelae of infectious and parasitic diseases, further subcategorised into viral and bacterial infection.

\begin{figure}[H]
\centering
\includegraphics[width=\textwidth]{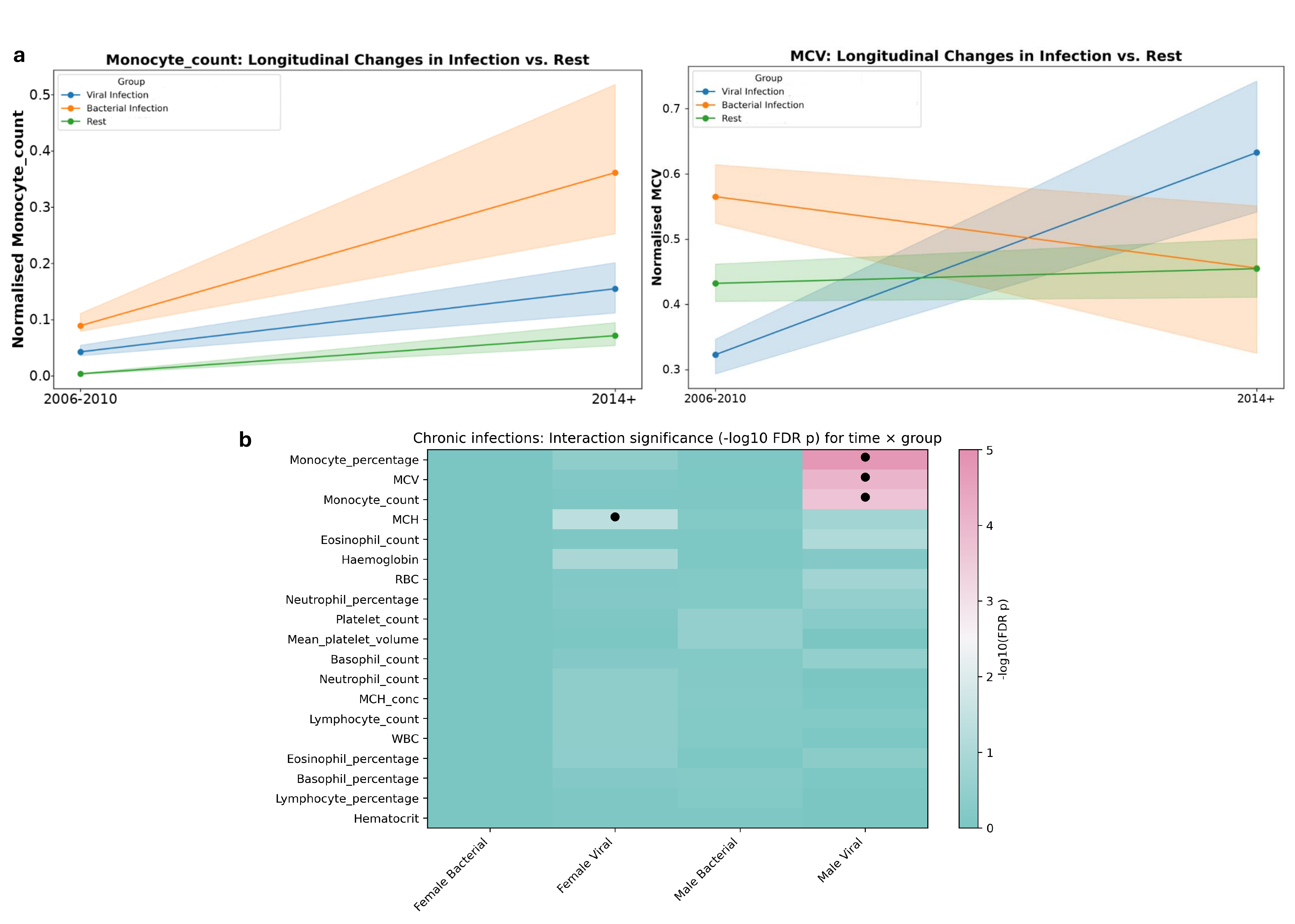}
\caption{(a) Median min-max normalised trajectories of representative CBC analytes across timepoints comparing chronic infection groups and Rest, with IQR shading. (b) Heatmap of time $\times$ infection interaction effects from linear mixed-effects (LME) models, shown as $-\log_{10}$ FDR-adjusted p-values; black dots indicate significant analytes (q $<$ 0.05).}
\label{fig:infection}
\end{figure}

Across incident chronic infection classifications, between-group comparisons of CBC markers (chronic viral infection, chronic bacterial infection, Rest) identified several significant pairwise differences after Bonferroni-corrected Mann–Whitney U testing, particularly for erythrocyte indices such as MCH and MCHC. However, these were not supported by significant global differences (Kruskal–Wallis), and no analytes met the full predefined criteria for robust between-group significance. Consistent with this, LME heatmap summaries showed only limited, heterogeneous signals across markers \autoref{fig:infection}. In contrast, longitudinal analysis revealed significant within-group changes in chronic viral infection, including decreases in MCH, platelet count, mean platelet volume, and monocyte count, with the largest effects in platelet-related markers. These trends are reflected in representative trajectories \autoref{fig:infection}, showing divergence from the relatively stable Rest group over time. No reliable longitudinal results were obtained for chronic bacterial infection due to limited sample size.

\section{Discussion and conclusion}

Distinguishing incident from prevalent disease proved central to our findings, as the former can offer a genuine window for early detection, while the latter is better suited to monitoring already-established disease. Restricting the index from disease-agnostic to disease type, and eventually subtype in cancer, further strengthened discrimination without requiring other disease-specific blood tests, and in some cases using only a limited number of core CBC analytes. Both distinctions shape our interpretation of the results. Longitudinal changes in routine blood biomarkers have been shown to precede cancer diagnosis by months to years, an underexploited signal this study seeks to leverage \cite{virdee2024}, while contrasting these trajectories across other major disease groups highlights potential sex- and disease-specific signatures. This is supported by large-scale genetic studies showing that common variation in haematological traits is biologically regulated and linked to complex disease pathways, reinforcing the relevance of CBC-derived signals beyond current clinical reliance on single-threshold measures \cite{soranzo2009genome} \cite{astle2016allelic}. This is consistent with research suggesting haematological traits reflect coordinated regulation of haematopoiesis across cell lineages, meaning longitudinal CBC changes capture dynamic shifts in these underlying processes \cite{vuckovic2020polygenic}.Analysis of normalised blood profiles revealed distinct biomarker patterns differentiating disease groups, suggesting that analyte selection combined with statistical and longitudinal approaches can uncover disease-specific, biologically meaningful immune signatures. Our overview of CVD, cancer, and chronic infection versus controls demonstrates clear longitudinal differences between groups, supporting the development of disease-agnostic machine-learning approaches to CBC and blood-based analysis, extending previous work \cite{john2025xgboost,leontiev2025exhaustive} and suggesting exploration of analyte-ratio-based indexes as a future direction.

Mean corpuscular haemoglobin (MCH), the average haemoglobin per cell \cite{zhang2022relationship}, was consistently lower across major disease groups, with downward trajectories in cancer and CVD. This aligns with prior reports linking reduced MCH to poorer survival in some cancers \cite{cui2020prognostic} and in CVD \cite{li2024nonlinear}. In cancer, iron dysfunction, specifically the hepcidin-ferroportin axis, has been hypothesised to impact disease development \cite{wu2014roles}; in CVD, MCH has been linked to macrocytosis, which is in turn associated with altered microcirculation and vascular damage \cite{li2024nonlinear}. Few longitudinal studies have examined MCH as a predictor of future disease in these categories, underscoring the novelty of our findings. In chronic infection, MCH abnormalities have been reported in both TB and HIV cohorts \cite{mwilitsa2024hematological, kibaru2015impact}, with longitudinal increases observed in HIV patients following antiretroviral therapy, suggesting dynamic erythropoietic changes during treatment \cite{kibaru2015impact}.

Longitudinal evidence on immune cell indices in CVD remains limited, but reduced pre-diagnosis monocyte levels in certain cancers have been linked to poor survival and disease progression \cite{edwards2023peripheral, rojko2020longitudinal}. Similar patterns appear in TB, where lower monocytes signal poor prognosis and higher pre-treatment levels predict better outcomes, and in HIV, where counts tend to normalise post-treatment \cite{shima2024monocytes, okeke2020longitudinal, mccausland2015altered}. Lower lymphocyte counts in CVD and cancer are consistent with existing literature, though longitudinal studies remain limited and largely focused on mortality or progression \cite{rojko2020longitudinal} \cite{phillips2011lymphocyte}. These patterns were supported by LME modelling of significant analytes \autoref{fig:cvd,cancer,infection}, suggesting a shared haematological suppression profile in CVD and cancer, contrasting with the immune-activated phenotype in chronic infection, highlighting the need for further granular analyses to identify disease-specific signatures within CVD and chronic infection in larger datasets.

Stratifying cancer into incident and prevalent cases, and further by subtype, reveals possible medication-related patterns and sustained alterations in blood analyte levels in prevalent disease, and early transitions from health to disease in incident cases. Moreover, heterogeneity within cancer cases is highlighted. Suppression of WBC and its differentials was observed in longitudinal analyses of both prevalent and incident cancer. WBC and their differentials have been widely reported to be reduced in individuals with previously diagnosed cancer \cite{cihan2013subtypes}, with chemotherapy contributing to depletion through its damaging effects on non-cancerous cells \cite{mackall1994lymphocyte} \cite{dixon2024sustained} \cite{kassie2025differences}. In incident cases, research suggests that abnormal monocyte counts may be indicative of developing cancer \cite{sajadieh2011monocyte} and could represent a time-dependent trajectory \cite{nost2021systemic} \cite{kresovich2020prediagnostic} consistent with our results. With respect to erythroid indices, MCH has been associated with several types of cancer incidence and mortality \cite{adris2018prospective}, acting as an independent risk factor \cite{shen2024hemoglobin}, with some studies reporting lower levels in prevalent cases as a consequence of chemotherapy-induced bone marrow suppression and chronic inflammation \cite{chinedu2025assessment}.Examining immune markers in incident blood and solid tumour cancer cases, lower pretreatment lymphocyte counts have been associated with poorer prognostic outcomes in solid tumours \cite{zhao2020prognostic}; however, limited research to date has explored longitudinal differences in blood biomarkers between cancer subtypes prior to diagnosis, highlighting the novelty of our research and the consistency of the results conforming to the medical literature. 

Previous work by our group introduced a single numerical score that aggregates CBC measures to quantify the deviation from individual baseline health \cite{hernandez2026integrative}. Here, we extend this framework by demonstrating an application to predictive medicine: LME modelling was used to account for repeated measures and identify significant analytes associated with specific cancer groups, which were ranked by effect size and included in the composite immune score. This reduced set of disease-specific CBC analytes is capable of providing predictive insights even in the absence of other key information, such as subjective data (symptoms) and contextual data (lifestyle), that traditionally complement a clinical profile. More analytes were significantly associated with disease in males than in females, likely reflecting the higher number of incident cancer cases in males in the UK Biobank \cite{gan2025impact}. The immune score demonstrated clear separation between the cancer groups and the control, with some distinct distributions observed between cancer types, specifically solid tumour cancers, indicating heterogeneity in haematologic and possibly immune alterations.

We have shown that while increasing the number of analytes can improve discrimination between disease groups, this effect plateaus after the top-ranked features identified by LME modelling, indicating diminishing returns with additional markers, potentially due to the inclusion of less informative or confounding analytes. This is consistent with recent work showing that a small number of carefully selected routine CBC analytes, or their combinations, can capture the majority of the predictive signal without increasing feature complexity \cite{leontiev2025exhaustive}. The longitudinal trajectories of the immune score further revealed divergence prior to diagnosis, supporting its ability to capture early disease-associated changes. Future work will extend this analyte-specific immune score framework to other disease groups beyond cancer and explore the development of patient-specific CBC-based indices.

Breast and prostate cancer showed similar immune scores when using all significant analytes, sharing three: RBC, lymphocyte percentage, and haemoglobin. RBC profiles are altered in the presence of breast tumours and have been proposed as a disease-monitoring marker \cite{pereira2022red, obeagu2025revolutionizing}, while in prostate cancer, erythroid-related analytes (RBC, WBC) have been linked to disease risk and mortality, potentially via testosterone's role in stimulating erythropoiesis \cite{watts2020hematologic}. This points to a possible sex-hormonal axis linking erythroid changes to these malignancies, consistent with our findings.

This study has some limitations. The UK Biobank cohort shows notable loss to follow-up, reducing participants with complete longitudinal blood measurements. Additionally, only three blood collection time points were available, and limited sample size at the final visit required merging the latter two assessments for modelling. Future work will aim to leverage larger longitudinal datasets across a broader range of diseases to support more disease-agnostic approaches. While identified analytes show strong discriminatory potential, some biomarkers will still require context-specific clinical interpretation, and further expansion of disease-specific marker evaluation may enhance precision. Nonetheless, identifying generalisable haematologic signatures that distinguish major disease groups represents a novel, meaningful result warranting further validation.

In summary, this study demonstrates that trend analysis of routine blood results, both in raw time-series patterns and through composite indexes, can reveal distinct disease-specific signatures using a limited set of informative analytes, with potential for clinically accessible approaches to personalised early detection and improved disease stratification.

\section{Methods}

The dataset included 19,367 UK Biobank participants with longitudinal complete blood count (CBC) measurements \cite{sudlow2015}, taken at initial assessment (2006–2010) and two follow-up visits (first repeat: 2012–13; final: from 2014 onward). Given 90\% loss to follow-up at each time point, the two repeat assessments were collated into a single data point for analysis. 

\subsection{Data preprocessing and case-control definition}

\begin{figure}[H]
\centering
\includegraphics[width=1\textwidth]{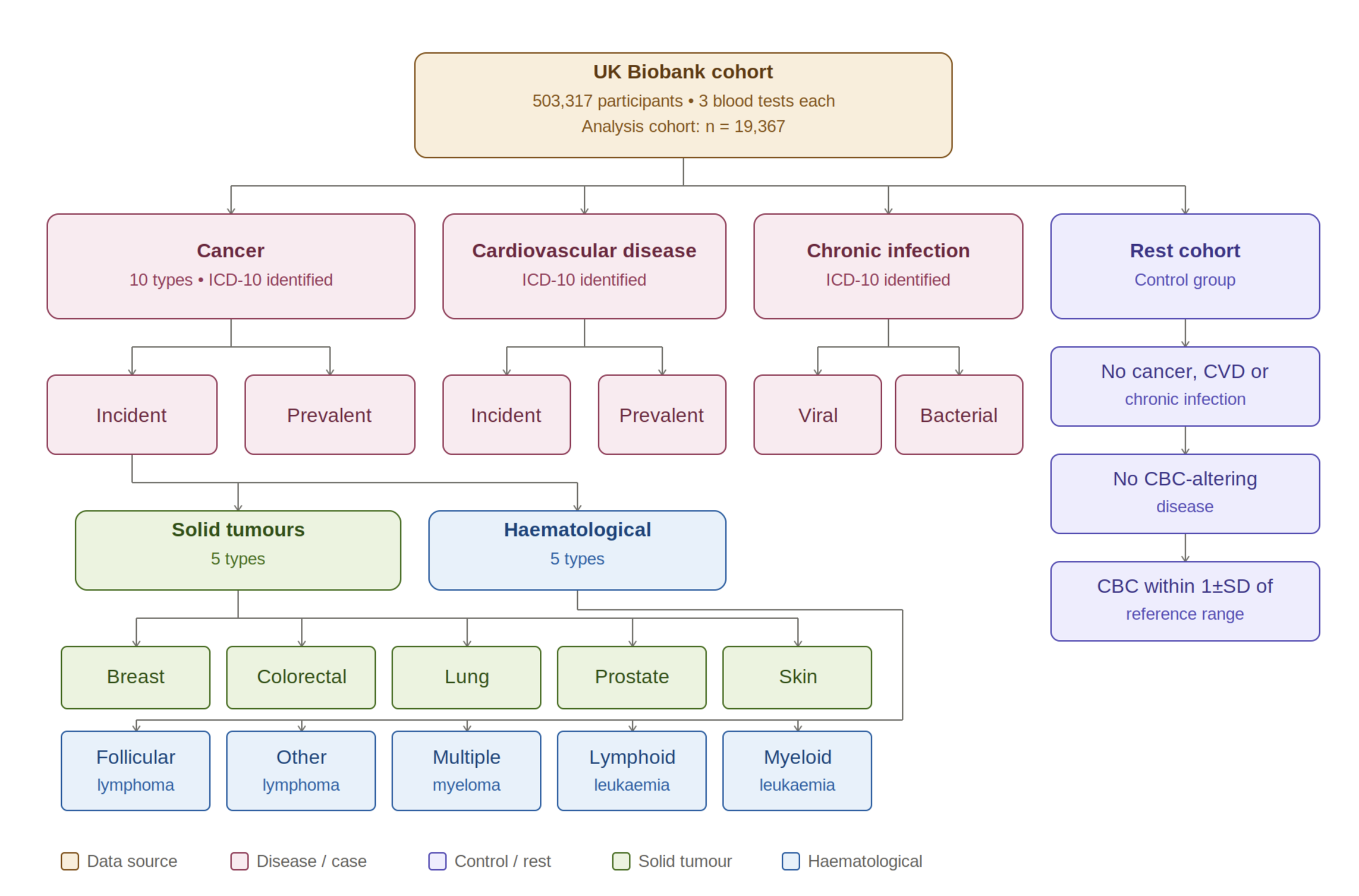}
\caption{Participant grouping by disease status and control classification. Only participants with at least two complete CBC measurements were included, and classified into major disease categories (Cancer, CVD, Chronic Infection) using ICD10 codes, diagnosis dates, and timing relative to baseline. Chronic infections were categorised as viral or bacterial; cancers were stratified into incident/prevalent and solid tumour/blood cancer, with further malignancy-specific subtypes.}
\label{fig:UKBBcohort}
\end{figure}

Data were preprocessed to include only participants with blood values from at least two assessments, then split into four categories: cancers, chronic infections, cardiovascular disease (CVD), and a "Rest" control group (\autoref{fig:UKBBcohort}). Diagnoses used main ICD10 codes and diagnosis dates. Controls had none of the diseases under investigation throughout the study, were not on medication significantly altering CBC values (e.g. blood thinners, immunosuppressants), and had initial blood values within $\pm$1 of NHS CBC reference ranges, allowing mild abnormalities; without this, no participants qualified as controls. Disease groups were split into prevalent (pre-recruitment) and incident (post-recruitment) cases to capture blood baseline changes across disease progression. Incident cancers were further divided into solid tumours and blood cancers, then subdivided into the cohort's most common malignancies (\autoref{fig:UKBBcohort}). CVD was split only into incident and prevalent cases, as overlapping diagnoses precluded mutually exclusive subgroups. Incident chronic infections were stratified into viral and bacterial categories, with no further sub-classification possible given the small number of UK Biobank cases.

Variance and distribution of all blood data were assessed using QQ plots and Shapiro–Wilk tests per biomarker, indicating non-normal distributions for most analytes. Primary analysis used Kruskal–Wallis tests to assess between-group differences (disease vs. control); where p p $\leq$ 0.05, post hoc pairwise Mann–Whitney U tests with Bonferroni correction were conducted, with Cohen's d calculated for effect size. Longitudinal within-group changes were assessed per analyte using Friedman tests followed by pairwise Wilcoxon tests, again with Cohen's d for effect size.

We employed median min-max normalised longitudinal plots per analyte with interquartile range (IQR) bands. Within-analyte normalisation enabled across-marker comparison, visualising longitudinal changes in disease groups relative to controls. Predictive value was assessed using progression divergence, where separation between IQR bands indicated distinct trajectories; trend direction, where consistent positive or negative trends indicated directional change even with overlapping IQRs; and rate of change, where slope steepness captured the degree of change over time, detectable even with overlapping IQRs.


We then applied a linear mixed-effects (LME) modelling framework to incident disease cases to characterise longitudinal haematological trajectories and identify analytes contributing to disease signatures, controlling for key individual-level covariates. Sex, age, and time point were included as fixed effects, with subject-specific random intercepts accounting for baseline inter-individual variability. Disease-by-time interaction terms were included per analyte to test whether trajectories differed by disease status, with significance evaluated using Benjamini--Hochberg FDR correction. Models were fitted separately by sex; analytes showing significant predictive associations---representing coordinated, multi-analyte temporal deviations from baseline ageing trajectories---were defined as disease-specific signatures, providing a quantitative framework to capture early, preclinical haematological alterations not apparent cross-sectionally. Interaction significance values were visualised as heatmaps, with colour intensity representing the strength of statistical evidence ($-\log_{10}$ FDR-adjusted p-value) for differential temporal dynamics between disease and control groups.

\subsection{Disease-specific dynamic expansion of immune indexes}
Significant analytes identified from the LME models were incorporated into the Immune Index framework, a previously published dimensional reduction method developed by our group to quantify blood-related markers for triaging purposes, also linked to biological age \cite{hernandez2026integrative}. The score aggregates numerical blood test results into a single value enabling rapid classification, triaging, and detection of deviation from personalised reference values; it is sometimes termed the ``immune'' index due to the CBC's connection to the immune system via red and white blood cells, though the CBC is best understood as the most scalable test closest to, rather than fully representative of, the broader immune system. The full mathematical description is in \cite{hernandez2026integrative}. The original framework used all available CBC analytes to construct a multidimensional immune reference space, drawing on NHANES and UK Biobank data. Here, beyond extending the original simulation-based framework to a longitudinal retrospective study, we restricted the immune space to disease-specific analyte subsets identified as significant by LME analysis. To assess their individual and cumulative contributions, analytes were added sequentially by significance, with resulting changes in immune score evaluated at each step using violin plots, quantifying the incremental impact of each analyte and whether additional features improved disease discrimination. Rather than computing deviations across the full CBC feature space, disease-specific immune scores were calculated within this reduced subspace, constructing disease-focused immune spaces while preserving the original distance-based scoring framework, and enabling immune deviation to be quantified using the most informative features for each disease.

\section{Code Availability} 

All code is provided in the Data and Availability section (See GitHub page). These intelligent systems were used to analyse the dataset, identify predictive features, that is, detect patterns and signatures modelling complex biological processes underlying disease states, and ultimately perform classification tasks for disease monitoring.

\section{Acknowledgements}

This research has been conducted using the UK Biobank Resource under Application Number 79893. 

\bibliographystyle{plainnat}
\bibliography{references}

\newpage

\section*{Supplementary Information}

\setcounter{figure}{0}
\renewcommand{\thefigure}{S\arabic{figure}}

\setcounter{table}{0}
\renewcommand{\thetable}{S\arabic{table}}

\begin{figure}[ht]
\centering
\includegraphics[width=1\textwidth]{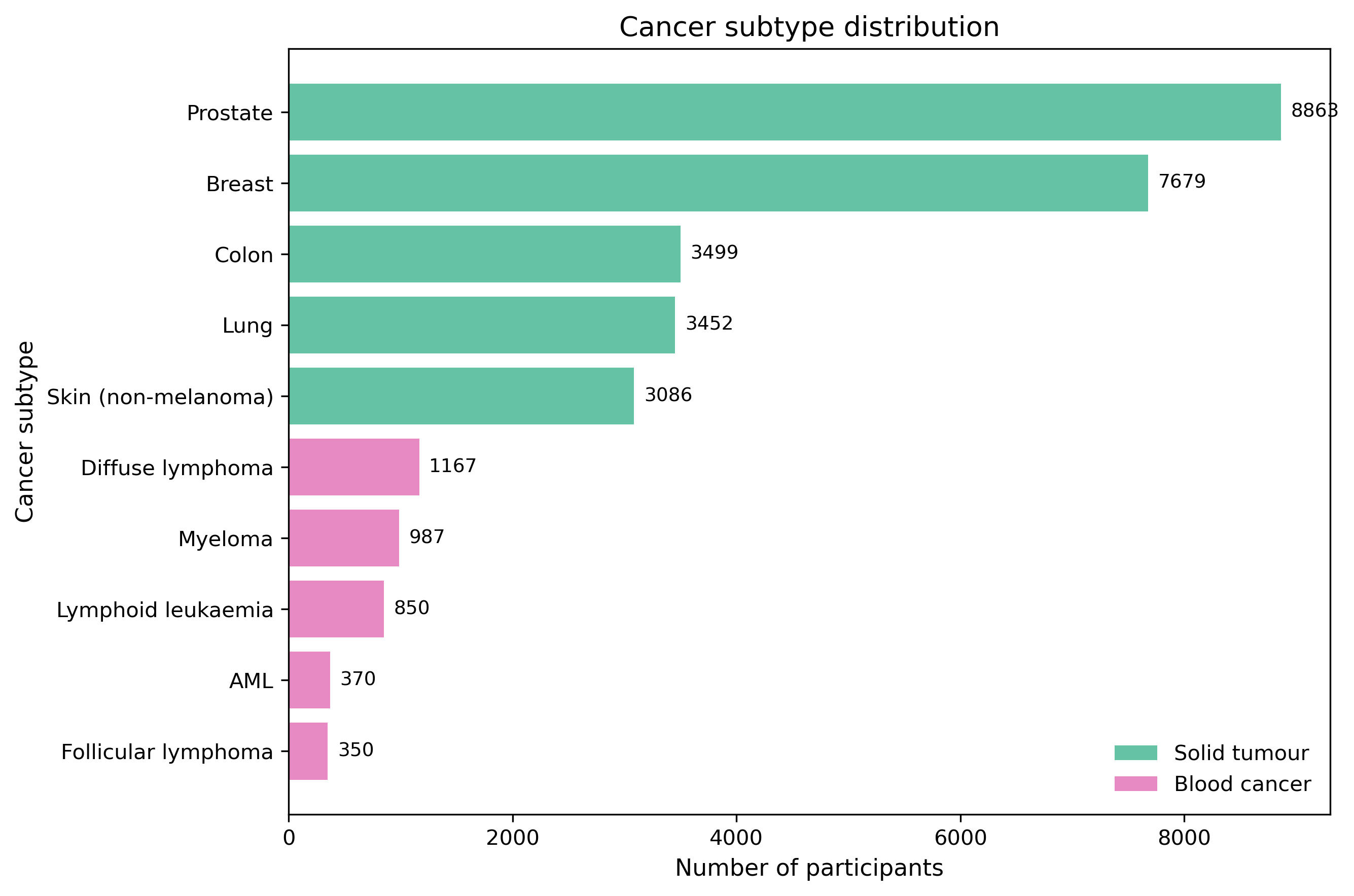}
\caption{Distribution of cancer subtypes in the study cohort. Cancer cases were stratified into subtype-specific groups and classified as solid tumours or blood cancers. Bar lengths indicate participant counts for each subtype.}
\label{fig:cancersubtypes}
\end{figure}

\begin{figure}[ht]
\centering
\includegraphics[width=1\textwidth]{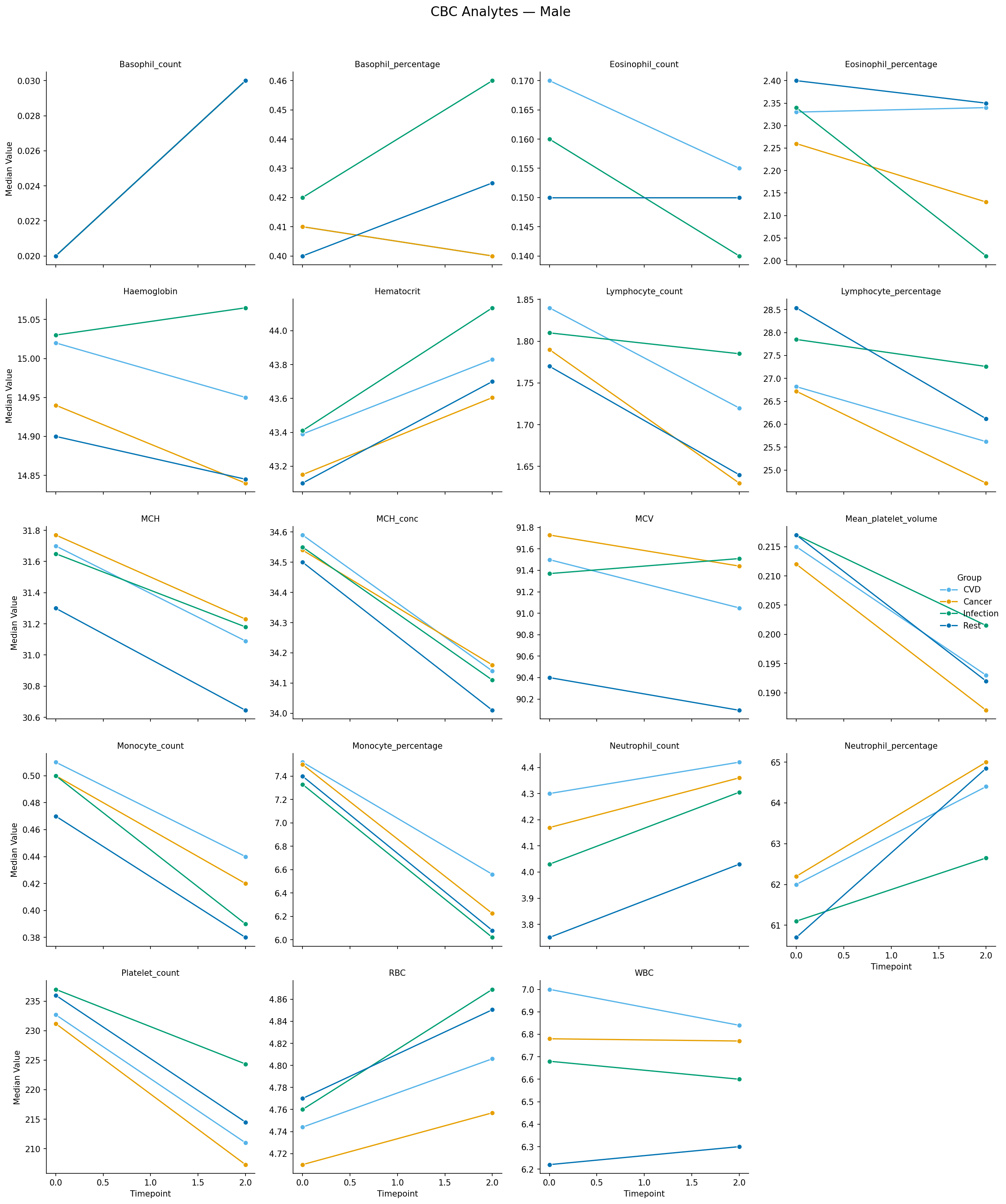}
\caption{Faceted line plots depicting the longitudinal trajectories of CBC analytes at three timepoints in male UK Biobank participants, grouped by disease status: cancer (yellow), cardiovascular disease/CVD (light blue), infection (green), and healthy controls (dark blue).}
\end{figure}

\begin{figure}[ht]
\centering
\includegraphics[width=1\textwidth]{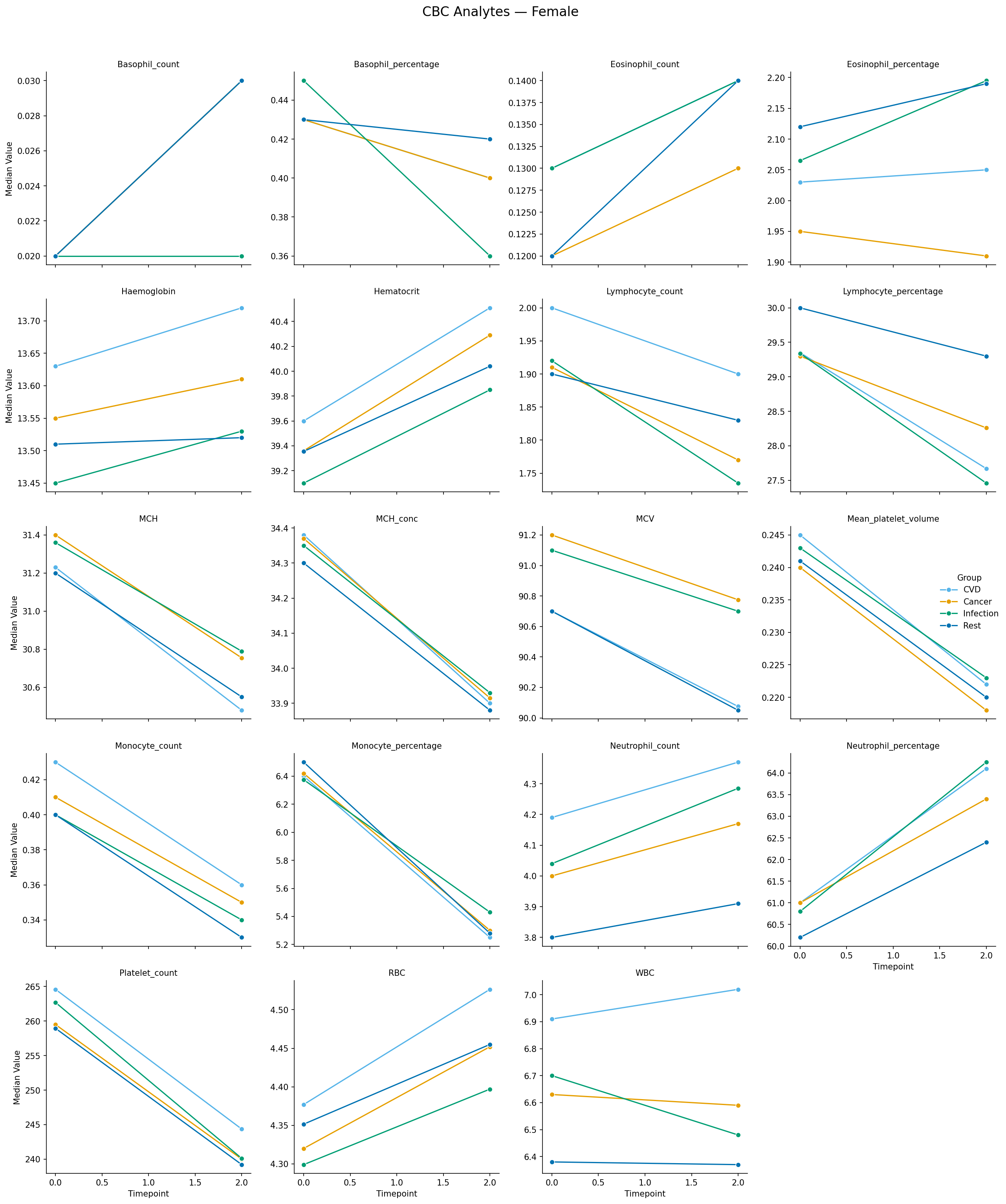}
\caption{Faceted line plots depicting the longitudinal trajectories of CBC analytes at three timepoints in female UK Biobank participants, grouped by disease status: cancer (yellow), cardiovascular disease/CVD (light blue), infection (green), and healthy controls (dark blue).}
\end{figure}

\begin{table}[ht]
\caption{Number of significant analytes per pairwise comparison, by timepoint.}
\label{tab:S1_pairwise_summary}
\begin{tabular}{llll}
\toprule
Comparison & Baseline & Follow-up & Delta \\
\midrule
CVD vs Cancer & 45 & 33 & 6 \\
CVD vs Rest & 44 & 39 & 18 \\
CVD vs Infection & 43 & 31 & 3 \\
Cancer vs Rest & 39 & 38 & 11 \\
Cancer vs Infection & 37 & 18 & 1 \\
Infection vs Rest & 37 & 10 & 1 \\
\bottomrule
\end{tabular}
\end{table}

\begin{table}
\caption{Top 20 analytes by absolute effect size.}
\label{tab:S2_top_effects}
\begin{tabular}{lllll}
\toprule
Analyte & Comparison & Timepoint & p (Bonferroni) & Cohen's d \\
\midrule
NRBC & Rest vs Infection & Follow-up & 3.74e-31 & -0.512 \\
Cholesterol & CVD vs Rest & Follow-up & 7.79e-127 & -0.512 \\
LDL\_direct & CVD vs Rest & Follow-up & 1.46e-119 & -0.493 \\
Urate & CVD vs Rest & Baseline & 8.07e-241 & 0.486 \\
Urate & CVD vs Infection & Baseline & $<1\times10^{-300}$ & 0.429 \\
Urate & CVD vs Rest & Follow-up & 1.81e-83 & 0.428 \\
Urate & CVD vs Infection & Follow-up & 4.15e-20 & 0.422 \\
Cystatin\_C & CVD vs Rest & Baseline & 3.97e-292 & 0.415 \\
WBC & CVD vs Rest & Baseline & 9.68e-182 & 0.401 \\
Cystatin\_C & CVD vs Rest & Follow-up & 5.14e-73 & 0.401 \\
NRBC & Rest vs Infection & Delta & 3.10e-21 & -0.399 \\
Apolipoprotein\_B & CVD vs Rest & Follow-up & 8.84e-81 & -0.394 \\
Cholesterol & CVD vs Infection & Follow-up & 5.60e-16 & -0.372 \\
Cholesterol & CVD vs Rest & Baseline & 6.10e-153 & -0.370 \\
Oestradiol & CVD vs Infection & Follow-up & 2.49e-4 & -0.366 \\
Neutrophil\_count & CVD vs Rest & Baseline & 1.71e-151 & 0.362 \\
LDL\_direct & CVD vs Rest & Baseline & 2.58e-144 & -0.357 \\
Cystatin\_C & Cancer vs Rest & Baseline & 2.06e-179 & 0.351 \\
MCH & Rest vs Infection & Follow-up & 1.18e-11 & -0.350 \\
HDL\_cholesterol & CVD vs Infection & Follow-up & 5.07e-13 & -0.349 \\
\bottomrule
\end{tabular}
\end{table}

\begin{table}
\caption{MCH, monocyte count and lymphocyte count — significant pairwise contrasts.}
\label{tab:S3_targeted_analytes}
\begin{tabular}{lllll}
\toprule
Analyte & Comparison & Timepoint & p\_bonferroni & cohens\_d \\
\midrule
Lymphocyte\_count & CVD vs Rest & Baseline & 7.50e-27 & 0.160 \\
Lymphocyte\_count & Infection vs Rest & Baseline & 2.10e-9 & 0.124 \\
Lymphocyte\_count & Cancer vs Rest & Baseline & 0.022 & 0.059 \\
Lymphocyte\_count & CVD vs Infection & Baseline & 2.15e-11 & 0.051 \\
Lymphocyte\_count & Cancer vs Infection & Baseline & 6.87e-11 & 0.022 \\
Lymphocyte\_count & CVD vs Cancer & Baseline & 1.69e-128 & -0.004 \\
Lymphocyte\_count & Rest vs Cancer & Follow-up & 3.38e-5 & -0.043 \\
Lymphocyte\_count & CVD vs Cancer & Follow-up & 0.002 & -0.033 \\
MCH & Cancer vs Rest & Baseline & 1.21e-78 & 0.219 \\
MCH & CVD vs Rest & Baseline & 3.21e-49 & 0.167 \\
MCH & Infection vs Rest & Baseline & 1.40e-39 & 0.160 \\
MCH & Cancer vs Infection & Baseline & 3.35e-11 & 0.063 \\
MCH & CVD vs Cancer & Baseline & 6.19e-28 & -0.057 \\
MCH & Rest vs Infection & Follow-up & 1.18e-11 & -0.350 \\
MCH & Rest vs Cancer & Follow-up & 2.32e-48 & -0.305 \\
MCH & CVD vs Rest & Follow-up & 7.72e-36 & 0.263 \\
Monocyte\_count & CVD vs Rest & Baseline & 4.22e-115 & 0.289 \\
Monocyte\_count & CVD vs Infection & Baseline & 3.59e-172 & 0.168 \\
Monocyte\_count & Infection vs Rest & Baseline & 4.33e-9 & 0.126 \\
Monocyte\_count & Cancer vs Rest & Baseline & 1.31e-32 & 0.098 \\
Monocyte\_count & CVD vs Cancer & Baseline & 1.95e-245 & 0.060 \\
Monocyte\_count & Cancer vs Infection & Baseline & 7.71e-20 & 0.039 \\
Monocyte\_count & CVD vs Rest & Follow-up & 1.79e-34 & 0.283 \\
Monocyte\_count & CVD vs Infection & Follow-up & 3.97e-7 & 0.200 \\
Monocyte\_count & Rest vs Cancer & Follow-up & 3.69e-13 & -0.169 \\
Monocyte\_count & Cancer vs Infection & Follow-up & 0.006 & 0.102 \\
Monocyte\_count & CVD vs Cancer & Follow-up & 6.82e-5 & 0.033 \\
\bottomrule
\end{tabular}
\end{table}

\begin{table}
\caption{CBC-only pairwise comparison summary.}
\label{tab:S4_CBC_only_summary}
\begin{tabular}{llll}
\toprule
Comparison & Baseline & Follow-up & Delta \\
\midrule
CVD vs Cancer & 16 & 11 & 0 \\
CVD vs Infection & 16 & 9 & 0 \\
CVD vs Rest & 16 & 14 & 5 \\
Cancer vs Infection & 15 & 5 & 0 \\
Cancer vs Rest & 15 & 15 & 4 \\
Infection vs Rest & 16 & 5 & 0 \\
\bottomrule
\end{tabular}
\end{table}

\begin{table}
\caption{Number of analytes with significant within-group change (Friedman p<0.05).}
\label{tab:S5_within_group_summary}
\begin{tabular}{ll}
\toprule
Group & n\_significant\_analytes \\
\midrule
Cancer & 19 \\
CVD & 17 \\
Rest & 16 \\
Infection & 11 \\
\bottomrule
\end{tabular}
\end{table}

{\small
\setlength{\tabcolsep}{7pt}
\begin{longtable}{
  l
  l
  c
  S[table-format=1.2e-2]
  S[table-format=-1.3]
  S[table-format=3.0]
}
\caption{Targeted within-group longitudinal effects for MCH, MCHC, mean platelet
volume (MPV), platelet count, basophil count and neutrophil percentage.
$p_{\text{adj}}$: Wilcoxon signed-rank test, Bonferroni-corrected.
$d$: Cohen's $d$ (positive values indicate an increase from the earlier to the later timepoint).}
\label{tab:S7_targeted_longitudinal} \\
\toprule
Analyte & Group & Comparison & {$p_{\text{adj}}$} & {Cohen's $d$} & {$n$} \\
\midrule
\endfirsthead
\caption[]{(continued)} \\
\toprule
Analyte & Group & Comparison & {$p_{\text{adj}}$} & {Cohen's $d$} & {$n$} \\
\midrule
\endhead
\bottomrule
\multicolumn{6}{r}{\footnotesize\emph{Continued on next page}} \\
\endfoot
\bottomrule
\endlastfoot
Basophil count & Rest      & T1--T2 & 2.24e-18 & -0.228 & 549 \\
               & Rest      & T0--T1 & 4.59e-5  &  0.164 & 549 \\
               & Cancer    & T1--T2 & 6.75e-9  & -0.133 & 420 \\
               & Rest      & T0--T2 & 2.02e-5  & -0.105 & 549 \\
               & CVD       & T1--T2 & 4.64e-4  & -0.046 & 479 \\
               & Cancer    & T0--T2 & 3.29e-5  &  0.006 & 420 \\
               & CVD       & T0--T2 & 9.00e-3  & -0.004 & 479 \\
\addlinespace
MCH            & Rest      & T1--T2 & 9.03e-70 &  1.024 & 549 \\
               & CVD       & T1--T2 & 3.97e-58 &  0.713 & 481 \\
               & Rest      & T0--T2 & 1.22e-49 &  0.701 & 549 \\
               & Cancer    & T1--T2 & 3.99e-47 &  0.617 & 423 \\
               & Cancer    & T0--T2 & 1.02e-34 &  0.514 & 423 \\
               & CVD       & T0--T2 & 3.26e-40 &  0.490 & 481 \\
               & Infection & T1--T2 & 1.99e-6  &  0.380 &  56 \\
               & Rest      & T0--T1 & 1.09e-5  & -0.196 & 549 \\
               & Infection & T0--T2 & 8.51e-6  &  0.194 &  56 \\
               & CVD       & T0--T1 & 9.13e-4  & -0.161 & 481 \\
\addlinespace
MCHC           & Rest      & T1--T2 & 3.18e-54 &  0.893 & 549 \\
               & CVD       & T1--T2 & 2.12e-46 &  0.855 & 481 \\
               & Cancer    & T1--T2 & 2.78e-42 &  0.782 & 423 \\
               & Infection & T1--T2 & 1.01e-6  &  0.681 &  56 \\
               & CVD       & T0--T2 & 2.13e-22 &  0.507 & 481 \\
               & Cancer    & T0--T2 & 8.24e-23 &  0.478 & 423 \\
               & Rest      & T0--T2 & 7.93e-21 &  0.465 & 549 \\
               & Rest      & T0--T1 & 9.89e-23 & -0.463 & 549 \\
               & CVD       & T0--T1 & 1.90e-17 & -0.403 & 481 \\
               & Cancer    & T0--T1 & 1.19e-11 & -0.334 & 423 \\
               & Infection & T0--T2 & 7.71e-6  &  0.331 &  56 \\
\addlinespace
MPV            & Rest      & T0--T2 & 1.17e-43 &  0.805 & 429 \\
               & Rest      & T0--T1 & 3.46e-44 &  0.793 & 429 \\
               & CVD       & T0--T2 & 3.56e-34 &  0.733 & 355 \\
               & CVD       & T0--T1 & 4.42e-34 &  0.702 & 355 \\
               & Cancer    & T0--T1 & 2.90e-27 &  0.568 & 342 \\
               & Cancer    & T0--T2 & 1.13e-23 &  0.565 & 342 \\
               & Infection & T0--T2 & 5.00e-3  &  0.504 &  42 \\
               & Infection & T0--T1 & 3.00e-2  &  0.343 &  42 \\
               & Rest      & T1--T2 & 4.10e-2  &  0.114 & 429 \\
\addlinespace
Neutrophil (\%) & Cancer   & T0--T2 & 1.40e-23 & -0.523 & 420 \\
               & CVD       & T0--T2 & 2.63e-12 & -0.356 & 479 \\
               & Cancer    & T1--T2 & 9.45e-12 & -0.318 & 420 \\
               & Rest      & T0--T2 & 7.08e-12 & -0.277 & 549 \\
               & CVD       & T1--T2 & 8.82e-7  & -0.231 & 479 \\
               & Cancer    & T0--T1 & 4.56e-4  & -0.199 & 420 \\
               & Rest      & T1--T2 & 1.74e-6  & -0.181 & 549 \\
               & CVD       & T0--T1 & 3.10e-2  & -0.121 & 479 \\
\addlinespace
Platelet count & Rest      & T0--T2 & 4.52e-46 &  0.691 & 549 \\
               & Rest      & T0--T1 & 5.07e-33 &  0.561 & 549 \\
               & CVD       & T0--T2 & 3.12e-31 &  0.547 & 481 \\
               & Cancer    & T0--T2 & 1.52e-23 &  0.498 & 423 \\
               & CVD       & T0--T1 & 1.02e-24 &  0.448 & 481 \\
               & Infection & T0--T2 & 3.98e-5  &  0.427 &  56 \\
               & Cancer    & T0--T1 & 5.01e-17 &  0.370 & 423 \\
               & Rest      & T1--T2 & 3.21e-8  &  0.217 & 549 \\
               & CVD       & T1--T2 & 7.61e-4  &  0.120 & 481 \\
               & Cancer    & T1--T2 & 1.50e-2  &  0.115 & 423 \\
\end{longtable}
}

\end{document}